\documentclass[traditabstract,epsf]{aa}
\usepackage{txfonts}
\usepackage{epsfig}

\newcommand{\teff}{\mbox{$T_{\rm eff}$}}
\newcommand{\logg}{\mbox{$\log g$ }}
\newcommand{\vsini}{\mbox{$v \sin{i_\ast}$}}
\newcommand{\mictrb}{\mbox{$\xi_{\rm t}$}}
\newcommand{\mactrb}{\mbox{$v_{\rm mac}$}}

\newcommand{\kms}{\mbox{km\,s$^{-1}$}}
\newcommand{\ms}{\mbox{m\,s$^{-1}$}}
\newcommand{\halpha}{\mbox{$H_\alpha$}}

\begin{document}
\title{WASP-50\,b: a hot Jupiter transiting a moderately active solar-type star\thanks{The photometric time-series used in this work are only available in electronic form at the CDS via anonymous ftp to  cdsarc.u-strasbg.fr (130.79.128.5) or via http://cdsweb.u-strasbg.fr/cgi-bin/qcat?J/A+A/}}

\author{ 
M.~Gillon\inst{1},
A.~P.~Doyle\inst{2},
M.~Lendl\inst{3},
P.~F.~L.~Maxted\inst{2}, 
A.~H.~M.~J.~Triaud\inst{3},
D.~R.~Anderson\inst{2}, 
S.~C.~C.~Barros\inst{4},
J.~Bento\inst{5}, 
A.~Collier-Cameron\inst{6},  
B.~Enoch\inst{6}, 
F.~Faedi\inst{4},
C.~Hellier\inst{2}, 
E.~Jehin\inst{1}, 
P.~Magain\inst{1}, 
J.~Montalb\'an\inst{1},  
F.~Pepe\inst{3},
D.~Pollacco\inst{4}, 
D.~Queloz\inst{3},
B.~Smalley\inst{2}, 
D.~Segransan\inst{3}, 
A.~M.~S.~Smith\inst{2}, 
J.~Southworth\inst{2},
S.~Udry\inst{3}, 
R.~G.~West\inst{7},
P.~J.~Wheatley\inst{5}
}

\offprints{michael.gillon@ulg.ac.be}
\institute{
$^1$ Universit\'e de Li\`ege, All\'ee du 6 ao\^ut 17, Sart Tilman, Li\`ege 1, Belgium \\
$^2$ Astrophysics Group, Keele University, Staffordshire, ST5 5BG, United Kingdom\\
$^3$ Observatoire de Gen\`eve, Universit\'e de Gen\`eve, 51 Chemin des Maillettes, 1290 Sauverny, Switzerland\\
$^4$ Astrophysics Research Centre, School of Mathematics \& Physics, Queen's University, University Road, Belfast, BT7 1NN, UK \\
$^5$ Department of Physics, University of Warwick, Coventry CV4 7AL, UK\\
$^6$ School of Physics and Astronomy, University of St. Andrews, North Haugh, Fife, KY16 9SS, UK\\
$^7$ Department of Physics and Astronomy, University of Leicester, Leicester, LE1 7RH, UK\\
}

\date{Received date / accepted date}
\authorrunning{M. Gillon et al.}
\titlerunning{A hot Jupiter transiting a moderately active solar-type star}
\abstract{
We report the discovery by the WASP transit survey of a giant planet in a close orbit 
($0.0295 \pm 0.0009$ AU) around a moderately bright ($V=11.6$, $K=10$) G9 dwarf 
($0.89 \pm 0.08$ $M_\odot$, $0.84 \pm 0.03$ $R_\odot$) in the Southern constellation Eridanus. 
Thanks to high-precision follow-up photometry and spectroscopy obtained by the telescopes 
TRAPPIST and {\it Euler}, the mass and size of this planet, WASP-50\,b, are well constrained 
to $1.47 \pm 0.09$ $M_{\rm Jup}$ and  $1.15 \pm 0.05$ $R_{\rm Jup}$, respectively. The transit ephemeris 
is $2455558.6120$ $(\pm 0.0002) + N \times 1.955096$ $(\pm 0.000005)$ HJD$_{\rm UTC}$.  The size of the planet is 
consistent with basic models of irradiated giant planets. The chromospheric activity ($\log R'_{HK} = -4.67$) 
and rotational period ($P_{\rm rot} = 16.3 \pm 0.5$~days) of the host star suggest an age of $0.8\pm 0.4$ Gy that is 
discrepant with a stellar-evolution estimate based on the measured stellar parameters 
($\rho_*  = 1.48 \pm 0.10$  $\rho_\odot $, \teff  = $5400\pm100$ K, [Fe/H]=$-0.12\pm0.08$) which favours an age of $7 \pm 3.5$ Gy. 
This discrepancy could be explained by the  tidal and magnetic influence of the planet on the star, 
in good agreement with the observations that stars hosting hot Jupiters tend to show faster rotation and 
magnetic activity (Pont 2009;  Hartman 2010). We measure
a stellar inclination of  $84_{-31}^{+6}$ deg, disfavouring a high stellar obliquity. 
Thanks to its large irradiation and the relatively small size of its host star,  
WASP-50\,b is a good target for occultation spectrophotometry, making it able to constrain  the 
relationship between hot Jupiters' atmospheric thermal profiles and the chromospheric  activity of 
their host stars proposed by Knutson et al. (2010). 
}

\maketitle

\section{Introduction}

While the {\it Kepler} space mission is pursuing its pioneering search for habitable terrestrial planets transiting solar-type stars (Borucki et al. 2011), ground-based, wide-field surveys continue to detect short-period (i.e. a few days, or even less) transiting giant planets at an increasing rate. These transiting `hot Jupiters' are important objects for the nascent field of comparative exoplanetology. As they could not form {\it in situ} at such short distances from their parent stars, their orbital configurations pose an interesting challenge for theories of formation and evolution of planetary systems (D'Angelo et al. 2010; Lubow \& Ida 2010). Their migrational history can be further constrained by the measurement of the orientation of their orbital axis relative to the stellar spin axis (Winn 2010a), the measurements obtained so far suggesting that past dynamical interactions with a third body could be driving their migration (Triaud et al. 2010). Hot Jupiters also make it possible to study the physical response of a giant planet to an irradiation orders of magnitude larger than in solar system planets (Burrows et al. 2008; Fortney et al. 2010), and also  to the strong gravitational and magnetic fields so close to their host stars (Correia \& Laskar 2010; Chang et al. 2010). Furthermore, these planets could be able to influence the properties of their parent stars, notably by  modifying their angular momentum budget and by inducing chromospheric activity (Lanza 2010). The atmospheres of hot Jupiters can also be studied thoroughly. Their generally large atmospheric scale heights and their frequent transits maximize the signal-to-noise ratios (SNR) achievable on short timescales in transit absorption spectrophotometry, while their large irradiation makes it possible to measure the thermal emission profile of their dayside through occultation-depth measurements at different wavelengths (Seager  2010). Such atmospheric measurements have made possible the study of the atmospheric composition of several planets (Seager \& Deming 2010). They have also revealed that most `hot Jupiters' harbour a high-altitude thermal inversion on their dayside, the presence or absence of this inversion being possibly correlated with the activity of the host star (Knutson et al. 2010). Furthermore, occultation measurements can help in measuring their orbital eccentricity, an important parameter to assess their energy budget and tidal history (Gillon et al. 2010a). These planets are thus real `exoplanetology laboratories', especially if they orbit stars bright enough to make possible high-SNR follow-up studies.

In that context, we announce here the discovery of a hot Jupiter, WASP-50\,b, that transits a moderately active,  $V$=11.6, G9V star in the Southern constellation Eridanus. Section~2 presents the discovery photometry obtained by the WASP transit survey, and  high-precision follow-up observations obtained from La Silla ESO Observatory by the TRAPPIST and {\it Euler} telescopes to confirm the transits and planetary nature of WASP-50\,b, and to determine precisely its parameters. In Sect.~3, we present the spectroscopic determination of the stellar properties and the derivation of the system parameters  through a combined analysis of the follow-up photometric and spectroscopic time-series. We discuss our results in Sect.~4.

\section{Observations}

\subsection{WASP photometry}

The host star WASP-50 (1SWASPJ025445.13--105353.0 = 2MASS02544513--1053530 = USNO-B1.0 0791-0028223 =  Tycho-2 5290-00462-1 = GSC1 5290-00462 = GSC2.3 S2L3000257; $V$=11.6, $K$=10.0) was observed by the Northern and Southern telescopes of the WASP survey (Pollacco et al. 2006) during the 2008 and 2009 observing seasons, covering the intervals 2008 Sep 16 to 2009 January 3 and  2009 August 30 to 2009 December 19. The 19731 pipeline-processed photometric measurements were de-trended and searched for transits using the methods described by Collier Cameron et al. (2006). The selection process (Collier Cameron et al. 2007) found WASP-50 as a high priority candidate showing a periodic transit-like signature with a period of 1.955~days. A total of 26 transits are observed in the data. Fig.~1 (top panel) presents the WASP photometry folded on the best-fitting transit ephemeris. 

We  analyzed the WASP lightcurve to determine whether it shows periodic
modulation due to the combination of  magnetic activity and the rotation of
the star.  The value of \vsini$ $ derived in section 3.1 together
with the estimated stellar radius implies a rotation period of 16\,--\,24 days,
assuming that the rotation axis of the star is approximately aligned with the
orbital axis of the planet. We subtracted the transit signal from the
data and then used the sine-wave fitting method described in Maxted et al.
(2011) to calculate the periodograms shown in Fig.~2. These are
calculated over 4096 uniformly spaced frequencies from 0 to 1.5 cycles/day.
The false alarm probability (FAP) levels shown in these figures are calculated using
a bootstrap Monte-Carlo method also described in Maxted et al. (2011).
Variability due to star spots is not expected to be coherent on long
timescales, as a consequence of the finite lifetime of star-spots and
differential rotation in the photosphere, so we analysed the two seasons of
data for WASP-50 separately. 

 Both seasons of data shows a significant periodic modulation with an
amplitude of 4\,--\,5\,milli-magnitudes, although at slightly different periods
---  $16.06\pm0.09$ days and  $16.65\pm0.13$ days. The FAP is
smaller than 0.05\% for  season 1, and $\sim$ 0.1\% for season 2. 
We used a bootstrap Monte-Carlo method to estimate the error on the periods. We also calculated periodograms of two stars with similar magnitudes 
and colours to WASP-50 observed simultaneously with the same camera. These stars 
show  significant periodic modulation at periods of 27 and 160 days and with amplitudes 
 of about 11\,milli-magnitudes in the first season of data. One of the stars shows
significant variability in the second season of data near a period of 1\,day,
the other shows no significant variability in this frequency range. Our
interpretation of these results is that WASP-50 does show a periodic
modulation of about 16.3\,days, but that instrumental noise in the WASP data
adds a systematic error of about $\pm 0.5$\,days to this estimate of the
rotation period. Combining the periods from both seasons and taking into 
account this systematic error, we conclude that the rotation period of WASP-50 is
 $16.3 \pm 0.5$\,days.  One could argue that the distribution of star spots on 
 the surface of a magnetically active star can produce periodic photometric 
 variations at half of the true rotation period. This is not very likely in the 
 case of WASP-50 because a star with a rotation period of 32 days is likely 
 to have very low levels of magnetic activity, and the implied value of \vsini$ $ 
 ($<1.3\,\kms$) would be inconsistent with the observed value given in Table 2.

\begin{figure}
\label{fig:1}
\centering                     
\includegraphics[width=9cm]{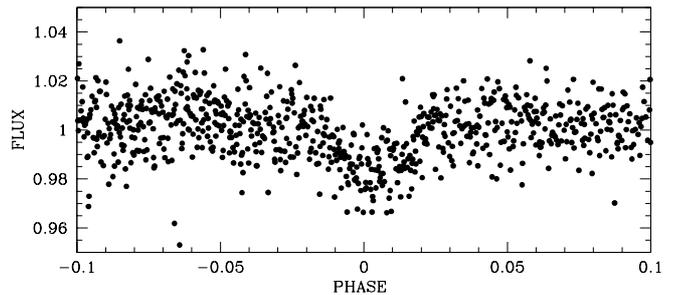}
\caption{WASP photometry of WASP-50 folded on the best-fitting transit ephemeris from 
the transit search algorithm presented in Collier Cameron et al. (2006), and binned per 5 measurements.} \end{figure}

\begin{figure*} 
\begin{center}
\includegraphics[width=0.9\textwidth]{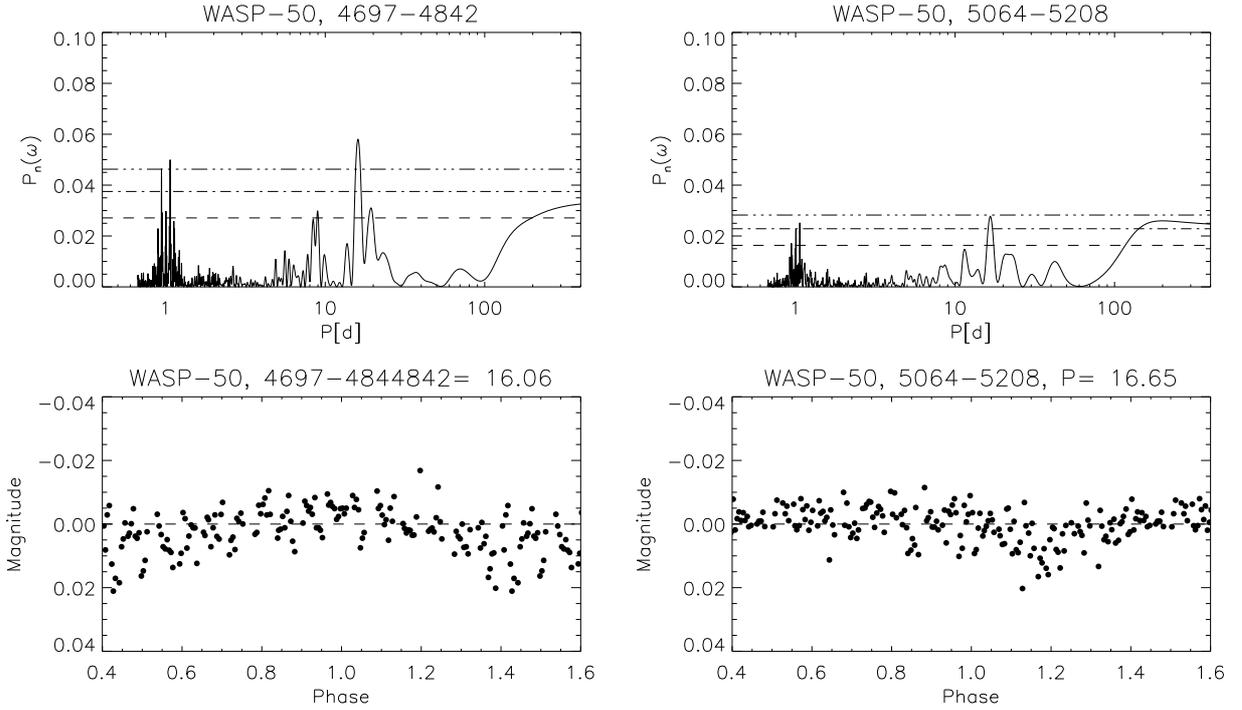}
\end{center}
\caption{{\it Upper panels}: periodograms for the WASP data from two seasons for
WASP-50. The date range (JD$-$245000) is given in the title of each panel.
Horizontal lines indicate false alarm probability levels FAP=0.1,0.01,0.001. 
{\it Lower panels}: data  folded and binned in 0.005 phase bins for the periods
noted together with the date range in the plot title.
 \label{swlomb} }
\end{figure*} 

\subsection{High-SNR transit photometry}

\subsubsection{TRAPPIST $I+z$-band photometry}

In order to confirm the origin of the transit signal on WASP-50 and to constrain thoroughly its parameters, three transits were observed at high-SNR with the new telescope TRAPPIST located at ESO La Silla Observatory in the Atacama Desert, Chile. TRAPPIST\footnote{see http://arachnos.astro.ulg.ac.be/Sci/Trappist} ({\it TRA}nsiting {\it P}lanets and {\it P}lanetes{\it I}mals {\it S}mall {\it T}elescope; Gillon et al. 2011) is a 60cm robotic telescope dedicated to high-precision photometry of exoplanet host stars and to the photometric monitoring of bright comets. It was installed at La Silla in April 2010. Its commissioning  ended in November 2010, and it has been operational since then. Its thermoelectrically-cooled camera is equipped with a 2K\,$\times$\,2K Fairchild 3041 CCD with a 22'\,$\times$\,22' field of view (scale = 0.65"/pixel). We observed three consecutive transits of WASP-50  on the nights of 2010 December 23, 25 and 27. We used the $2\times2$ MHz read-out mode with 1 $\times$ 1 binning, resulting in a typical readout + overhead time  and readout noise of 6.1\,s and 13.5\,$e^{-}$, respectively. A slight defocus was applied to the telescope to optimize the observation efficiency and to minimize pixel to pixel effects. We observed through a special `$I$+$z$' filter that has a transmittance $>$90\% from 750 nm to beyond 1100 nm. The positions of the stars on the chip were maintained to within a few pixels over the course of the three runs, thanks to the `software guiding' system that regularly derives  an astrometric solution to the most recently acquired image and sends pointing corrections to the mount if needed. 

Weather conditions were good during the three runs. The integration time was 25s for the first run and 20s for the other runs.  After a standard pre-reduction (bias, dark, flatfield correction), the stellar fluxes were extracted from the images using the {\tt IRAF/DAOPHOT}\footnote{{\tt IRAF} is distributed by the National Optical Astronomy Observatory, which is operated by the Association of Universities for Research in Astronomy, Inc., under cooperative agreement with the National Science Foundation.} aperture photometry software (Stetson, 1987). Several sets of reduction parameters were tested, and we kept the one giving the least scatter for stars of similar brightness to WASP-50. After a careful selection of reference stars, differential photometry was then obtained.  Fig.~3 shows the TRAPPIST transit light curves divided by the best-fitting baseline model, with the best-fitting transit model superimposed.

\begin{figure*}
\label{fig:trappist}
\centering                     
\includegraphics[width=18cm]{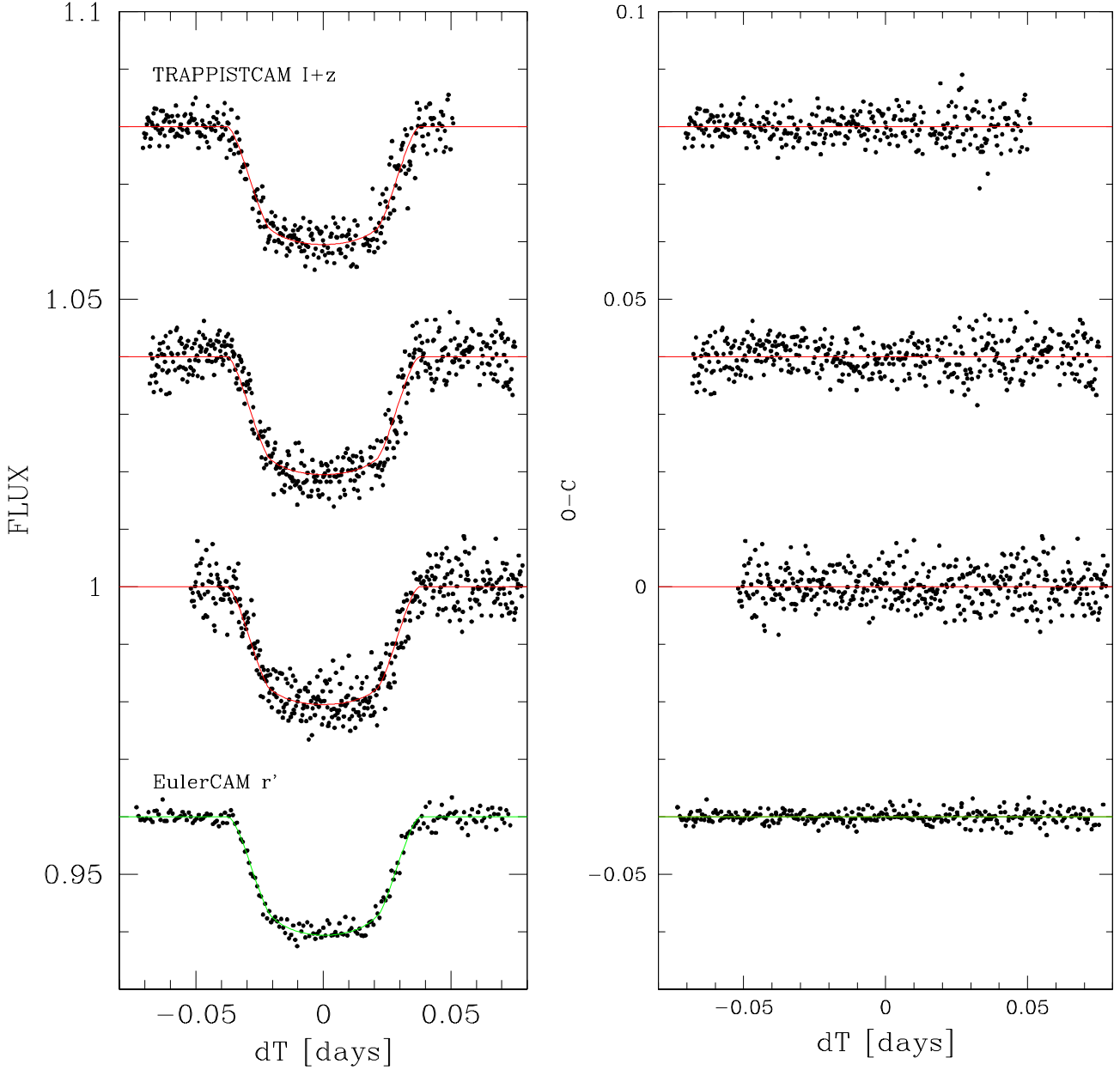}
\caption{{\it Left}, from $top$ to $bottom$: period-folded TRAPPIST transit light curves of WASP-50 for 2010 December 23, 25, and 27, and {\it Euler} transit light curve.  These light curves are divided by the best-fitting baseline models (see details in Sec.~4), and the best-fitting transit models are superimposed in red (TRAPPIST) and green ({\it Euler}). Three light curves and models are shifted in the $y$-axis for the sake of clarity. $Right$:  the residuals to the fit for each light curve.}
\end{figure*}

\subsubsection{{\it Euler} $r$-band photometry}

The transit of 25 Dec 2010 was also observed in the $r$\,Gunn filter with the EulerCam CCD camera at the 1.2-m {\it Euler} Telescope at La Silla Observatory. This nitrogen-cooled camera has a 4k\,$\times$\,4k E2V CCD with a 15'\,$\times$\,15'  field of view  (scale=0.23"/pixel). The exposure time was 40\,s, the readout + overhead time being 25\,s (readout noise $\sim$\,4.5\,$e^{-}$). Here too, a defocus was applied to the telescope to optimize the observation efficiency and minimize pixel to pixel effects, while flat-field effects were further reduced by keeping the stars on the same pixels, thanks to a `software guiding' system similar to TRAPPIST's. However, a residual drift of the target on the detector was observed for airmasses larger than 1.5. This positional drift did not result in any clear drift in the photometry itself. Furthermore, any potential systematic effect on the photometric baseline is included in our analysis under the form of a second-order time polynomial (see Sect.~3.2). The reduction was similar to the one of the TRAPPIST data, i.e. aperture photometry was extracted for all stars in the pre-reduced images, then differential photometry was obtained.  The resulting light curve is shown on Fig.~3 (bottom). 

\subsection{Spectroscopy and radial velocities}

Once WASP-50 was identified as a high priority candidate we gathered spectroscopic measurements with the {\tt CORALIE} spectrograph mounted on {\it Euler} to confirm the planetary nature of the transiting body and obtain a mass measurement. 15 spectra were obtained from 2010 December 5 to 2011 January 5 with an exposure time of 30 minutes. Radial velocities  (RV)  were computed by weighted cross-correlation (Baranne et al. 1996) with a numerical G2-spectral template (Table~1). 

The RV time-series shows variations that are  consistent with a planetary-mass companion. A preliminary orbital analysis of the RVs resulted in a period and phase in excellent agreement with those deduced from the WASP transit detections (Fig.~4).  Assuming a stellar mass $M_{\ast} = 0.91 \pm 0.07$ $M_{\sun}$ (Sect.~3), the fitted semi-amplitude $K = 260 \pm 5$ m\,s$^{-1}$ translates into a planetary mass slightly larger than Jupiter's, $M_p = 1.52 \pm 0.08$ $M_{\rm Jup}$. The resulting orbital eccentricity is consistent with zero, $e = 0.01_{-0.01}^{+0.02}$. The scatter of the residuals is 10.8 \ms, very close to the average RV error, 10.5 \ms.

To exclude the possibility that the RV signal originates from spectral line distortions caused by a blended eclipsing binary, we analyzed the {\tt CORALIE} cross-correlation functions using the line-bisector technique described in Queloz et al (2001). The bisector spans seem quite stable, their standard deviation being nearly equal to their average error (22 $vs$ 21 \ms). No evidence for a correlation between the RVs and the bisector spans was found (Fig.~5), the deduced correlation coefficient being --0.24. 
 Considering not only this absence of correlation, but also the excellent agreement between the stellar densities inferred from the high-SNR transit photometry (Sect.~3.2) and from the spectroscopic analysis (Sect.~3.1), we concluded that a giant planet transits WASP-50 every $\sim$1.96 days.

\begin{figure}
\label{fig:rv}
\centering                     
\includegraphics[width=8.5cm]{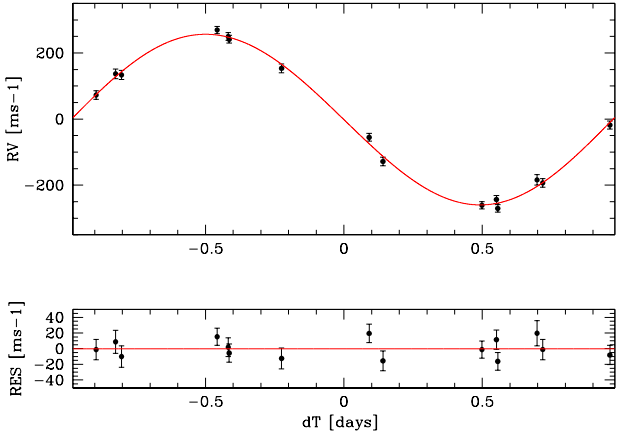}
\caption{$Top$: Euler/{\tt CORALIE} RVs with the best-fitting Keplerian model superimposed in red. $Bottom$: residuals of the fit.}
\end{figure}

\begin{figure}
\label{fig:rv}
\centering                     
\includegraphics[width=8.5cm]{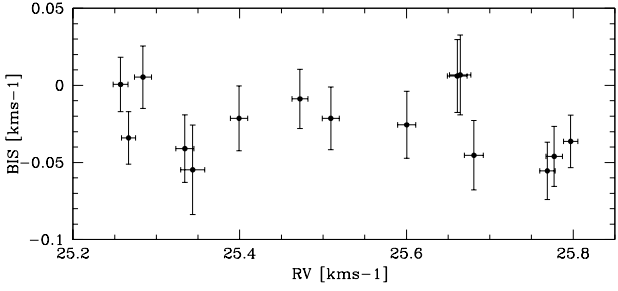}
\caption{Bisector versus RV measured from the {\tt CORALIE} spectra. We adopt uncertainties of twice the RV uncertainty for all bisector measurements. There is no correlation between these two parameters indicating the RV variations are not caused by stellar activity or line-of-sight binarity.}
\end{figure}

\begin{table}
\begin{center}
\begin{tabular}{cccc}
\hline
HJD-2,450,000 & RV & $\sigma_{RV}$ & BS\\ 
 & (km~s$^{-1}$) & (m~s$^{-1}$) & (km~s$^{-1}$)\\ \hline \noalign {\smallskip} 
 5535.647447 & 25.2667 &  8.5 & --0.0341  \\ \noalign {\smallskip} 
 5536.647197 & 25.7970 &  8.5 & --0.0363  \\ \noalign {\smallskip} 
 5537.656077 & 25.2840 & 10.1 & 0.0054   \\ \noalign {\smallskip} 
 5538.646866 & 25.7690 &  9.3 & --0.0554   \\ \noalign {\smallskip} 
 5541.573212 & 25.2571 &  8.8 & 0.0006    \\ \noalign {\smallskip} 
 5541.735749 & 25.3342 & 11.0 & --0.0410  \\ \noalign {\smallskip} 
 5542.556058 & 25.7772 &  9.7 & --0.0461   \\ \noalign {\smallskip} 
 5542.748534 & 25.6809 & 11.3 & --0.0453  \\ \noalign {\smallskip} 
 5543.671498 & 25.3436 & 14.5 & --0.0548  \\ \noalign {\smallskip} 
 5557.628117 & 25.5093 & 10.1 & --0.0214  \\ \noalign {\smallskip} 
 5561.641322 & 25.6004 & 10.9 & --0.0256  \\ \noalign {\smallskip} 
 5562.627319 & 25.4723 &  9.6  & --0.0088  \\ \noalign {\smallskip} 
 5563.667964 & 25.6645 & 13.0 &  0.0068   \\ \noalign {\smallskip} 
 5564.633499 & 25.3992 & 10.5 & --0.0214  \\ \noalign {\smallskip} 
 5565.645557 & 25.6610 & 11.8 &  0.0061   \\ \noalign {\smallskip} 
\hline
\end{tabular}
\caption{{\tt CORALIE} radial-velocity measurements for WASP-50 (BS = bisector spans).}
\end{center}
\label{wasp50-rvs}
\end{table}

\section{Analysis}

\subsection{Spectroscopic analysis - stellar properties}

The {\tt CORALIE} spectra of WASP-50 were co-added to produce a
single spectrum with a typical SNR of 100. The standard pipeline
reduction products were used in the analysis.

The analysis was performed using the methods given by
Gillon et al. (2009a). The \halpha\ line was used to determine the
effective temperature (\teff), while the Na\,{\sc i}\,D and Mg\,{\sc i}\,b lines
were used as surface gravity ($\log g$) diagnostics. The parameters 
obtained from the analysis are listed in Table~2. 
The elemental abundances were determined from equivalent width measurements 
of several clean and unblended lines. A value for microturbulence (\mictrb) 
was determined from Fe~{\sc i} lines using the method of Magain (1984). The 
quoted error estimates include those given by the uncertainties in \teff, \logg\ 
and \mictrb, as well as the scatter due to measurement and atomic data uncertainties.

The projected stellar rotation velocity (\vsini) was determined by fitting the
profiles of several unblended Fe~{\sc i} lines. A value for macroturbulence
(\mactrb) of 2.1 $\pm$ 0.3 {\kms} was assumed, based on the tabulation by
Gray (2008). An instrumental FWHM of 0.11 $\pm$ 0.01~{\AA} was
determined from the telluric lines around 6300\AA. A best-fitting value of
\vsini\ = 2.6 $\pm$ 0.5 \kms\ was obtained.

\begin{table}[h]
\begin{center}
\begin{tabular}{cc} \hline  
Parameter  & Value \\ \hline 
\teff      &   5400 $\pm$ 100 K \\
\logg      &   4.5 $\pm$ 0.1 \\
\mictrb    &   0.8 $\pm$ 0.2 \kms \\
\vsini     &   2.6 $\pm$ 0.5 \kms \\
{[Fe/H]}   &$-$0.12 $\pm$ 0.08 \\
{[Na/H]}   &   0.00 $\pm$ 0.10 \\ 
{[Mg/H]}   &$-$0.06 $\pm$ 0.03 \\
{[Si/H]}   &$-$0.01 $\pm$ 0.09 \\
{[Ca/H]}   &$-$0.04 $\pm$ 0.08 \\
{[Ti/H]}   &   0.00 $\pm$ 0.08 \\ 
{[V/H]}    &   0.03 $\pm$ 0.12 \\
{[Cr/H]}   &   0.00 $\pm$ 0.08 \\ 
{[Co/H]}   &   0.02 $\pm$ 0.04 \\
{[Ni/H]}   &$-$0.07 $\pm$ 0.06 \\
log A(Li)  &   $<$ 0.6 $\pm$ 0.2 \\
Mass       &   0.91 $\pm$ 0.07 $M_{\sun}$ \\
Radius     &   0.88 $\pm$ 0.11 $R_{\sun}$ \\
Sp. Type   &   G9 \\
Distance   &   230 $\pm$ 40 pc \\ \hline 
\\
\end{tabular}
\end{center}
\label{wasp50-params}
\caption{Stellar parameters of WASP-50 from spectroscopic analysis.
\newline
{\bf Note:} Mass and Radius estimate using the
calibration of Torres et al. (2010). Spectral type estimated from \teff\
using the table in Gray (2008).}
\end{table}

There is no significant detection of lithium in the spectra, with an equivalent
width upper limit of 3m\AA, corresponding to an abundance upper limit of
log~A(Li) $<$ 0.6 $\pm$ 0.2. The lack of lithium suggests an age $\ga 0.6$ Gyr,
 since similar levels of lithium depletion are found for stars of the same
spectral types in the Hyades and Praesepe clusters (Sestito \& Randich, 2005).

The rotation rate ($P_{\rm rot} = 16.85 \pm 3.7$~d) implied by the {\vsini} 
(assuming $\sin{i_\ast}=1$) is in excellent agreement with the period measured
from the WASP photometry, $16.3 \pm 0.5$~days. This latter implies a young
system, the deduced age being $\sim 0.8 \pm 0.4$~Gy using the
Barnes (2007) gyrochronology relation. 

The presence of Ca H+K emission (Fig.~6) indicates that WASP-50 is an active star, 
the deduced spectral index $\log R'_{HK}$ being $\sim -4.67$\footnote{We note that a transformation
for  $\log R'_{HK}$ measured with {\tt CORALIE} spectra to a standard system does not yet exist.}.
This value is similar to that found for the magnetically active star WASP-41 (Maxted et al. 2011). Thus
the spectroscopic activity indicates that the observed photometric variability of the star (Sect. 2.1) 
is due to the presence of spots.

\begin{figure}
\label{fig:5}
\centering                     
\includegraphics[angle=270,width=9cm]{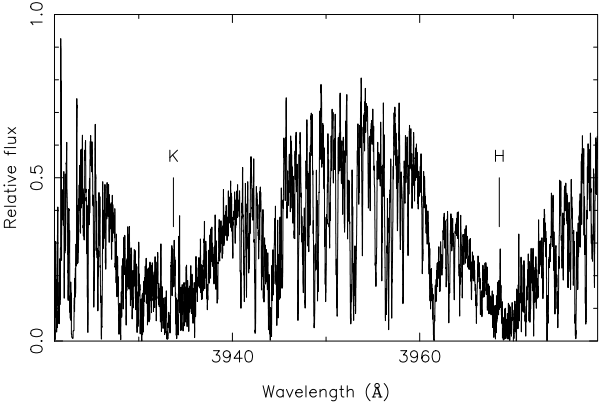}
\caption{A region of the co-added {\tt CORALIE} spectrum of WASP-50 showing emission in
the core of the CaII H and K lines.}
\end{figure}

\subsection{Global analysis}

We performed a global analysis of the follow-up photometry and the {\tt CORALIE}
 RV measurements. The transit timing deduced from the analysis of the 
WASP photometry, $2454997.4996 \pm 0.0014$ HJD\footnote{All the HJD presented 
in this work are based on the UTC time standard.}, was also used as input data to better
constrain the orbital period. The analysis was performed using the adaptive 
MCMC algorithm described in detail by Gillon et al. (2009b; 2010a; 2010b). To summarize, 
we simultaneously fitted the data, using for the photometry the transit model of Mandel 
\& Agol (2002) multiplied  by a baseline model consisting of a different second-order 
time polynomial for each of the four light curves, and  a Keplerian orbital 
model for the RVs.

The jump parameters\footnote{Jump parameters are the parameters that are randomly 
perturbed at each step of the MCMC.} in our MCMC analysis were: the planet/star area 
ratio $(R_p /R_\ast )^2$, the transit width (from first to last contact) $W$,  the parameter $b' = 
a \cos{i_p}/R_\ast$ (which is the transit impact parameter in case of a circular orbit),  the orbital 
period $P$ and time of minimum light $T_0$, the two parameters $\sqrt{e} \cos{\omega}$ 
and $\sqrt{e} \sin{\omega}$ where $e$ is the orbital eccentricity and $\omega$ is the argument 
of periastron, and the parameter $K_2 = K  \sqrt{1-e^2}   \textrm{ } P^{1/3}$, where $K$ is the 
RV orbital semi-amplitude. We assumed a uniform prior distribution for all these jump parameters. 
The photometric baseline model parameters and the systemic radial velocity of the star were 
not actual jump parameters, they were determined by least-square minimization at each step of
the MCMC.

We assumed a quadratic limb-darkening law, and we allowed  the quadratic coefficients $u_1$ 
and $u_2$ to float in our MCMC analysis, using as jump parameters not these coefficients 
themselves but the combinations $c_1 = 2 \times u_1 + u_2$  and $c_2 = u_1 - 2 \times u_2$ 
to minimize the correlation of the uncertainties (Holman et al. 2006).  To obtain a 
limb-darkening solution consistent with theory, we decided to use  normal prior distributions 
for $u_1$ and $u_2$ based on theoretical values and 1-$\sigma$ errors interpolated in the 
tables by Claret (2000; 2004): $u_1 = 0.29 \pm 0.035$ and $u_2 = 0.29 \pm 0.015$ for the $I+z$ 
filter, and $u_1 = 0.43 \pm 0.025$ and $u_2 = 0.28 \pm 0.015$ for the $r$ filter.

Our analysis was composed of five Markov chains of $10^5$ steps, the first 20\% of each 
chain being considered as its burn-in phase and discarded. For each run the convergence of the 
five Markov chains was checked using the statistical test presented by Gelman and Rubin (1992). 
The correlated noise present in the light curves was taken into account as described by Gillon et al. (2009b). 
For the RVs a  `jitter' noise of 7.5 \ms\  was added quadratically to the error bars, to equalize the 
mean error with the $rms$ of the best-fitting model residuals, 13 \ms. {This jitter level is compatible with the activity level of the star (Wright 2005).}

At each step of the Markov chains the stellar density deduced from the jump parameters, 
and values for  $T_{\rm eff}$ and  [Fe/H] drawn from the normal distributions deduced from our 
spectroscopic analysis (Sect.~3.1), were used as input for a modified version of the stellar-mass
 calibration law deduced by Torres et al. (2010) from well-constrained detached binary systems 
 (see Appendix A for details). Using the resulting stellar mass the physical parameters of the system 
 were then deduced from the jump parameters at each MCMC step. 
 
 Our measurement of the rotational period  of the star $P_{\rm rot}$ made it possible to derive the posterior 
 distribution for the stellar inclination $i_\ast$. As in the analysis of Watson et al. (2010), at each step of the Markov chains we used the value of the stellar radius $R_\ast$, and 
 values for $P_{\rm rot}$ and \vsini drawn from the normal distributions $N(16.3,0.5^2)$ days and 
 $N(2.6,0.5^2)$ \kms, to derive a value for $\sin{i_\ast}$ from
 
 \begin{equation}
 \sin{i_\ast} = \frac{P_{\rm rot} v\sin{i_\ast}}{2 \pi R_\ast}
  \end{equation} For $\sin{i_\ast} > 1$, $\sin{i_\ast}$ was set to 1 and the inclination $i_\ast$ was set to 
  90 deg. 
   
The parameters derived for the WASP-50 system can be found in Table 2. To assess the reliability
of the deduced physical parameters we performed a stellar evolution modeling based on the code 
CLES (Scuflaire et al. 2008), using as input the stellar density deduced from
our MCMC analysis and the effective temperature and metallicity deduced from our spectroscopic
analysis. The resulting  stellar mass was $0.84 \pm 0.05$ $M_\odot$, in good agreement with the 
MCMC result,  $0.89 \pm 0.08$ $M_\odot$. This stellar modeling led to an age of 
$7.0 \pm 3.5$ Gyr for the system. 

Fig.~7 shows the marginalized posterior distributions for the planet mass, radius, orbital inclination 
and eccentricity. As shown by the narrowness of these distributions, the physical parameters of the planet 
are well constrained, thanks to the precise determination of the transit and orbital parameters
made possible by the high-SNR of our follow-up data (transit photometry and RVs). In particular, 
the orbit appears to be close to circular, any eccentricity greater than 0.05 being firmly discarded. 

\begin{figure}
\label{fig:mr}
\centering                     
\includegraphics[width=8.5cm]{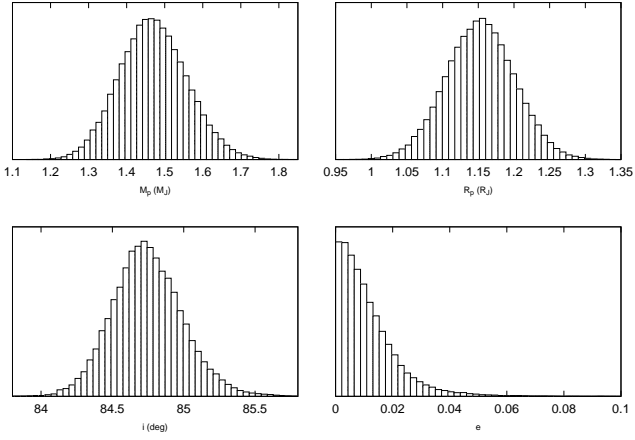}
\caption{Marginalized posterior distributions for the planet mass, radius, orbital inclination and eccentricity
resulting from our global MCMC analysis.}
\end{figure}

\begin{table}[h]
\begin{center}
\begin{tabular}{cc}
\hline \noalign {\smallskip}
$Jump$ $parameters$ &  \\ \noalign {\smallskip}
\hline \noalign {\smallskip}
Planet/star area ratio  $ (R_p/R_\ast)^2 $           &  $0.01970 \pm 0.00035$                          \\ \noalign {\smallskip} 
$b'=a\cos{i_p}/R_\ast$ [$R_*$]                            &  $0.689_{-0.018}^{+0.016}$                      \\ \noalign {\smallskip} 
Transit width  $W$ [d]                                          &   $0.07524 \pm 0.00068$                          \\ \noalign {\smallskip} 
$T_0-2450000$ [HJD]                                          &  $5558.61197_{-0.00015}^{+0.00021}$    \\ \noalign {\smallskip}
Orbital period  $ P$ [d]                                         &  $1.9550959 \pm 0.0000051$                    \\ \noalign {\smallskip} 
RV $K_2$  [m\,s$^{-1}$\,d$^{1/3}$]                      & $320.9 \pm 5.5$                                        \\ \noalign {\smallskip} 
$\sqrt{e}\cos{\omega}$                                         &  $0.042_{-0.064}^{+0.054}$                      \\ \noalign {\smallskip} 
$\sqrt{e}\sin{\omega}$                                          &  $0.038_{-0.079}^{+0.056}$                      \\ \noalign {\smallskip} 
$c1_{r'}$                                                               & $1.102 \pm 0.044$                                    \\ \noalign {\smallskip}  
$c2_{r'}$                                                               & $-0.132 \pm 0.039$                                   \\ \noalign {\smallskip}                             
$c1_{I+z}$                                                            & $0.865 \pm 0.062$                                     \\ \noalign {\smallskip}                            
$c2_{I+z}$                                                            & $-0.295 \pm 0.045$                                    \\ \noalign {\smallskip}                            
\hline \noalign {\smallskip}
$Deduced$ $stellar$ $parameters$   &    \\ \noalign {\smallskip}
\hline \noalign {\smallskip}
$u1_{r'}$                                                             & $0.414 \pm 0.022$                 \\ \noalign {\smallskip} 
$u2_{r'}$                                                             & $0.273 \pm 0.015$                 \\ \noalign {\smallskip} 
$u1_{I+z}$                                                          & $0.287 \pm 0.031$                 \\ \noalign {\smallskip} 
$u2_{I+z}$                                                          & $0.291 \pm 0.015$                 \\ \noalign {\smallskip} 
Density $\rho_* $  [$\rho_\odot $]                       & $1.48_{-0.09}^{+0.10}$          \\ \noalign {\smallskip} 
Surface gravity $\log g_*$ [cgs]                          & $4.537 \pm 0.022$                 \\ \noalign {\smallskip} 
Mass $M_\ast $    [$M_\odot$]                            & $0.892_{-0.074}^{+0.080}$    \\ \noalign {\smallskip} 
Radius  $ R_\ast $   [$R_\odot$]                         & $0.843 \pm 0.031$                 \\ \noalign {\smallskip} 
$\sin i_\ast$                                                         & $0.99_{-0.20}^{+0.01}$                     \\ \noalign {\smallskip} 
$i_\ast$ [deg]                                                       & $84_{-31}^{+6}$                     \\ \noalign {\smallskip} \hline \noalign {\smallskip}
$Deduced$ $planet$ $parameters$   &    \\ \noalign {\smallskip}
\hline \noalign {\smallskip}
RV $K$ [\ms]                                                        & $256.6 \pm 4.4$                              \\ \noalign {\smallskip} 
$R_p/R_\ast$                                                       & $0.1404 \pm 0.0013$                      \\ \noalign {\smallskip} 
$b_{tr}$ [$R_\ast$]                                               & $0.687_{-0.016}^{+0.014}$             \\ \noalign {\smallskip} 
$b_{oc}$ [$R_\ast$]                                             & $0.691_{-0.022}^{+0.019}$             \\ \noalign {\smallskip} 
$T_{oc}-2450000$ [HJD]                                     &  $5561.549 \pm 0.010$                    \\ \noalign {\smallskip} 
Orbital semi-major axis $ a $ [AU]                       &  $0.02945 \pm 0.00085$                  \\ \noalign {\smallskip} 
$a / R_\ast$                                                          & $7.51_{-0.15}^{+0.17}$                     \\ \noalign {\smallskip} 
Orbital inclination $i_p$ [deg]                               & $84.74 \pm 0.24$                            \\ \noalign {\smallskip} 
Orbital eccentricity $ e $                                       & $0.009_{-0.006}^{+0.011}$             \\ \noalign {\smallskip}
Argument of periastron  $ \omega $ [deg]            & $44_{-80}^{+62}$                            \\ \noalign {\smallskip} 
Equilibrium temperature $T_{eq}$ [K]$ $$^a$      & $1393 \pm 30$                                \\ \noalign  {\smallskip} 
Density  $ \rho_p$ [$\rho_{\rm Jup}$]                         &  $0.958_{-0.082}^{+0.095}$              \\ \noalign {\smallskip} 
Surface gravity $\log g_p$ [cgs]                           &  $3.439 \pm 0.025$                          \\ \noalign  {\smallskip} 
Mass  $ M_p$ [$M_{\rm Jup}$]                                   & $1.468_{-0.086}^{+0.091}$              \\ \noalign {\smallskip} 
Radius  $ R_p $ [$R_{\rm Jup}$]                                & $1.153 \pm 0.048$                           \\ \noalign {\smallskip} 
\hline \noalign {\smallskip}
\end{tabular}
\caption{Median and 1-$\sigma$ limits of the posterior distributions obtained for the WASP-50 
system derived from our MCMC analysis. $^a$Assuming $A$=0 and $F$=1. }
\end{center}
\end{table}

\section{Discussion}

WASP-50\,b is a new planet slightly larger ($1.15 \pm 0.05$ $R_{\rm Jup}$) and more massive ($1.47 \pm 0.09$ $M_{\rm Jup}$)  than Jupiter, in a close orbit ($0.0295 \pm 0.0009$ AU) around a moderately active G9V star ($0.89 \pm 0.08 M_\odot$, $0.84 \pm 0.03 R_\odot$) at $230 \pm 40$ pc from the solar system. Taking into account the large irradiation received by the planet ($8.7 \pm 1 \times 10^8$ erg\,s$^{-1}$\,cm$^{-2}$), the measured planetary mass and size are in good agreement with basic models for  irradiated planets (Fortney et al. 2007). Using these models our measured planet mass and radius suggest that the core of WASP-50b represents only a few per cent, or at most 10 per cent of its total mass, depending on whether we take the age as $\sim$1 Gy, as suggested by the rotation period of the star, or as older, as suggested by our stellar evolution modeling.  Fig.~8 shows the location of WASP-50b in a mass--radius diagram, showing only planets with masses between 1 and 2 $M_{\rm Jup}$. In this mass range the most extreme planets are the very dense CoRoT-13\,b (Cabrera et al. 2010), and the highly bloated WASP-12\,b (Hebb et al. 2009; Chan et al. 2011). WASP-50\,b appears to be a much more `normal' planet with a density close to Jupiter's. We note however that the extremely high density of CoRoT-13\,b was recently questioned by Southworth (2011).

With a size larger than Jupiter's, a high irradiation, and a relatively small and infrared-bright ($K$=10) host star, WASP-50\,b is a good target for  near-IR thermal emission measurements with the occultation photometry technique, using {\it Warm Spitzer} at 3.6 and 4.5 $\mu$m (e.g. Deming et al. 2011) or ground-based instruments at shorter wavelengths (e.g. Gillon et al. 2009b).  WASP-50\,b orbits an active star (as does, for example, WASP-41\,b, Maxted et al. 2010), making it able to test the proposition by Knutson et al. (2010) that the presence of thermal emission in the day-side atmosphere of a hot Jupiter is dependent of the chromospheric activity of its host star. This hypothesis suggests that an optical high-opacity source is present at high altitude in the dayside atmosphere of hot Jupiters and causes a thermal inversion, but that it is destroyed by the large UV flux produced by chromospherically active stars. This hypothesis is supported by emission measurements obtained so far, but it is still based on a small sample of systems. 

The values deduced in this work for $\log R'_{HK}$ and the planet density $\rho_p$ are consistent with the correlation between these two parameters noticed by Hartman (2010). As mentioned above, stellar evolution models and the measured density and spectroscopic parameters of WASP-50 point to an age of $7 \pm 3.5$ Gy, translating into an expected rotational period of $50 \pm 15$ days under the empirical relationship of Barnes (2007), calibrated on a large sample of cluster and field stars. Our measured rotation period ($16.3 \pm 0.5$ days) is in poor agreement with this expectation and suggests a younger age. In the context of a $\log R'_{HK}$-$\log g_p$ correlation, this discrepancy could be explained by the action of the planet itself on the star. In this scheme, hot Jupiters orbiting moderately cool stars like WASP-50 would tidally spin-up the stellar convective envelope and tilt its rotation axis, the stellar magnetic braking preventing the system from reaching spin--orbit synchronization, in agreement with the observations that stars hosting a hot Jupiter tend to rotate faster (Pont 2009), and that hot Jupiters with a high stellar obliquity are preferentially found around hot stars having no convective envelope (Winn et al. 2010b). Moreover, we note that our deduced stellar inclination $i_\ast$  ($84_{-31}^{+6}$  deg), although imprecise due to uncertainty in the \vsini $ $ determination, argues against a large obliquity. 

\begin{figure}
\label{fig:mr}
\centering                     
\includegraphics[width=8.5cm]{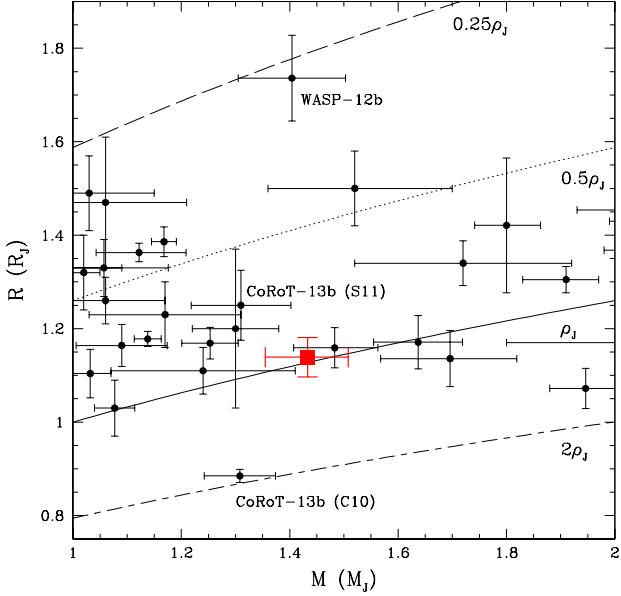}
\caption{Mass--radius diagram for the transiting planets with masses
ranging from 1 to 2 $M_{\rm Jup}$. WASP-50\,b is shown as a red square symbol. Two sets of values are shown for the planet CoRoT-13\,b, the ones
presented in the discovery paper (Cabrera et al. 2010, C10) and the ones recently reported by Southworth (2011, S11).}
\end{figure}

\begin{acknowledgements}
WASP-South is hosted by the South African Astronomical Observatory and we are grateful for their ongoing support and assistance. Funding for WASP comes from consortium universities and from UK's Science and Technology Facilities Council. TRAPPIST is a project funded by the Belgian Fund for Scientific Research (Fond National de la Recherche Scientifique, F.R.S-FNRS) under grant FRFC 2.5.594.09.F, with the participation of the Swiss National Science Fundation (SNF).  M. Gillon and E. Jehin are FNRS Research Associates. 
\end{acknowledgements} 

\bibliographystyle{aa}

\appendix
\section{Stellar calibration law}

Using a large sample of well constrained detached binaries, Torres et al. (2010; hereafter T10) 
showed that accurate stellar masses and radii could be deduced from an empirical law 
relating these two parameters to the  measured spectroscopic parameters \teff, \logg, 
and [Fe/H]. Enoch et al. (2010) showed that this law could be advantageously used in 
the analysis of transiting planets data with the MCMC algorithm, provided that the 
stellar surface gravity was replaced by the stellar density in the calibration law. Indeed, 
the stellar density is directly and precisely constrained by high-quality transit photometry, 
making it possible to obtain better stellar parameters than by using the stellar surface 
gravity deduced from a spectroscopic analysis (Winn 2010a). Under this condition, the 
calibration law makes it possible to deduce the mass of the star, and thus the physical parameters of 
the system, at each step of the MCMC, without the need for a concomitant or subsequent
stellar evolution modeling. The main limitation of this approach is that the T10 law is poorly constrained for stars below 0.6 $M_\odot$ and for pre-main-sequence stars.

The calibration law of Enoch et al. (2010) relating $M_\ast$ to $\rho_\ast$, $\teff$ and [Fe/H] is
\begin{eqnarray}
\log M_\ast  &=  &  a_1 + a_2 X + a_3 X^2 + a_4 (\log \rho_\ast) + \nonumber\\
& & a_5 (\log \rho_\ast)^2 +  a_6 (\log \rho_\ast)^3 + a_7 \textrm{[Fe/H], }
\end{eqnarray}  where $X$ is $\log \teff - 4.1$. 

We performed a linear regression analysis with a Singular Value Decomposition
algorithm (Press et al. 1992) to derive the best-fit values for the $a_{1:7}$  coefficients, using as data
the tabulated values presented by T10 for 19 binary systems for which the metallicity is known.
We added another very well characterized star to these data: the Sun, assuming 
conservatively a relative error of $10^{-3}$ on its mass and radius, 0.01 dex on its metallicity and 
 10 K on its effective temperature (5777 K, Cox 2000). 1-$\sigma$ errors were derived
 through a Monte Carlo analysis. The fitted and error values for the $a_{1:7}$  coefficients
 are shown in Table A.1. The resulting scatter in fitted-minus-measured values is $0.084$ $M_\odot$ 
 (0.023 in $\log M_\ast$).
 
To take into account the internal uncertainties of the calibration law, the $a_{1:7}$  coefficients
of Eq. A.1 are not fixed during our MCMC analysis but are drawn randomly from the normal distributions
shown in Table A.1 at each step of the Markov chains.

\begin{table}
\begin{center}
\caption{Coefficients for stellar mass calibration law.}
\begin{tabular}{cc}
\hline \noalign {\smallskip}
$a_1$  &   $0.450 \pm 0.013$     \\ \noalign {\smallskip} 
$a_2$  &  $1.421 \pm 0.057$       \\ \noalign {\smallskip} 
$a_3$  &  $0.33 \pm 0.19$         \\ \noalign {\smallskip} 
$a_4$  &  $-0.082 \pm 0.016$     \\ \noalign {\smallskip} 
$a_5$  & $0.032 \pm 0.017$        \\ \noalign {\smallskip} 
$a_6$  & $0.0026 \pm 0.0037$   \\ \noalign {\smallskip} 
$a_7$  & $0.075 \pm 0.012$       \\ \noalign {\smallskip} 
\hline
\end{tabular}
\end{center}
\end{table}
\end{document}